\documentclass{aa}
\usepackage{upgreek}
\usepackage{graphicx}
\usepackage[varg]{txfonts}
\usepackage{natbib}
\usepackage{float}
\usepackage{soul}
\newlength{\linwx}
\usepackage{color}

\begin{document}

\title{Dry or water world? How the water contents of inner sub-Neptunes constrain giant planet formation and the location of the water ice line}

\author{
Bertram Bitsch \inst{1}, Sean N. Raymond \inst{2}, Lars A. Buchhave \inst{3}, Aaron Bello-Arufe \inst{3}, Alexander D. Rathcke \inst{3}, and Aaron David Schneider\inst{4,5}
}
\offprints{B. Bitsch,\\ \email{bitsch@mpia.de}}
\institute{Max-Planck-Institut f\"ur Astronomie, K\"onigstuhl 17, 69117 Heidelberg, Germany
 \and
 Laboratoire d'Astrophysique de Bordeaux, CNRS and Universit\'e de Bordeaux, All\'ee Geoffroy St. Hilaire, 33165 Pessac, France
 \and
  National Space Institute, Technical University of Denmark, Elektrovej, 2800, Kgs. Lyngby, Denmark.
  \and
  Niels Bohr Institutet, Københavns Universitet, Blegdamsvej 17, 2100 København, Denmark
  \and
  Instituut voor Sterrenkunde, KU Leuven, 200 D, bus 2401, Celestijnenlaan 3001, 3001 Leuven, Belgium
}
\abstract{In the pebble accretion scenario, the pebbles that form planets drift inward from the outer disk regions, carrying water ice with them. At the water ice line, the water ice on the inward drifting pebbles evaporates and is released into the gas phase, resulting in water-rich gas and dry pebbles that move into the inner disk regions. Large planetary cores can block the inward drifting pebbles by forming a pressure bump outside their orbit in the protoplanetary disk. Depending on the relative position of a growing planetary core relative to the water ice line, water-rich pebbles might be blocked outside or inside the water ice line. Pebbles blocked outside the water ice line do not evaporate and thus do not release their water vapor into the gas phase, resulting in a dry inner disk, while pebbles blocked inside the water ice line release their water vapor into the gas phase, resulting in water vapor diffusing into the inner disk. As a consequence, close-in sub-Neptunes that accrete some gas from the disk should be dry or wet, respectively, if outer gas giants are outside  or inside the water ice line, assuming that giant planets form fast, as has been suggested for Jupiter in our Solar System. Alternatively, a sub-Neptune could form outside the water ice line, accreting a large amount of icy pebbles and then migrating inward as a very wet sub-Neptune. We suggest that the water content of inner sub-Neptunes in systems with giant planets that can efficiently block the inward drifting pebbles could constrain the formation conditions of these systems, thus making these sub-Neptunes exciting targets for detailed characterization (e.g., with JWST, ELT, or ARIEL). In addition, the search for giant planets in systems with already characterized sub-Neptunes can be used to constrain the formation conditions of giant planets as well.
}
\authorrunning{Bitsch et al.}\titlerunning{Dry or water world?}\maketitle

\section{Introduction}
\label{sec:Introduction}

Close-in transiting sub-Neptunes, planets of several Earth masses with atmospheres, can be characterized via observations, and it can be determined if the planetary atmosphere contains significant amounts of water \citep{2019ApJ...887L..14B, 2019NatAs...3.1086T, 2020arXiv200607444K}. Today, only a handful of planets are characterized in this detailed way, but the number of characterized sub-Neptunes is expected to increase significantly with the {\it James Webb Space Telescope} (JWST) and with the future ARIEL mission as well as with the ELT.

\begin{figure*}
 \centering
 \includegraphics[scale=0.133]{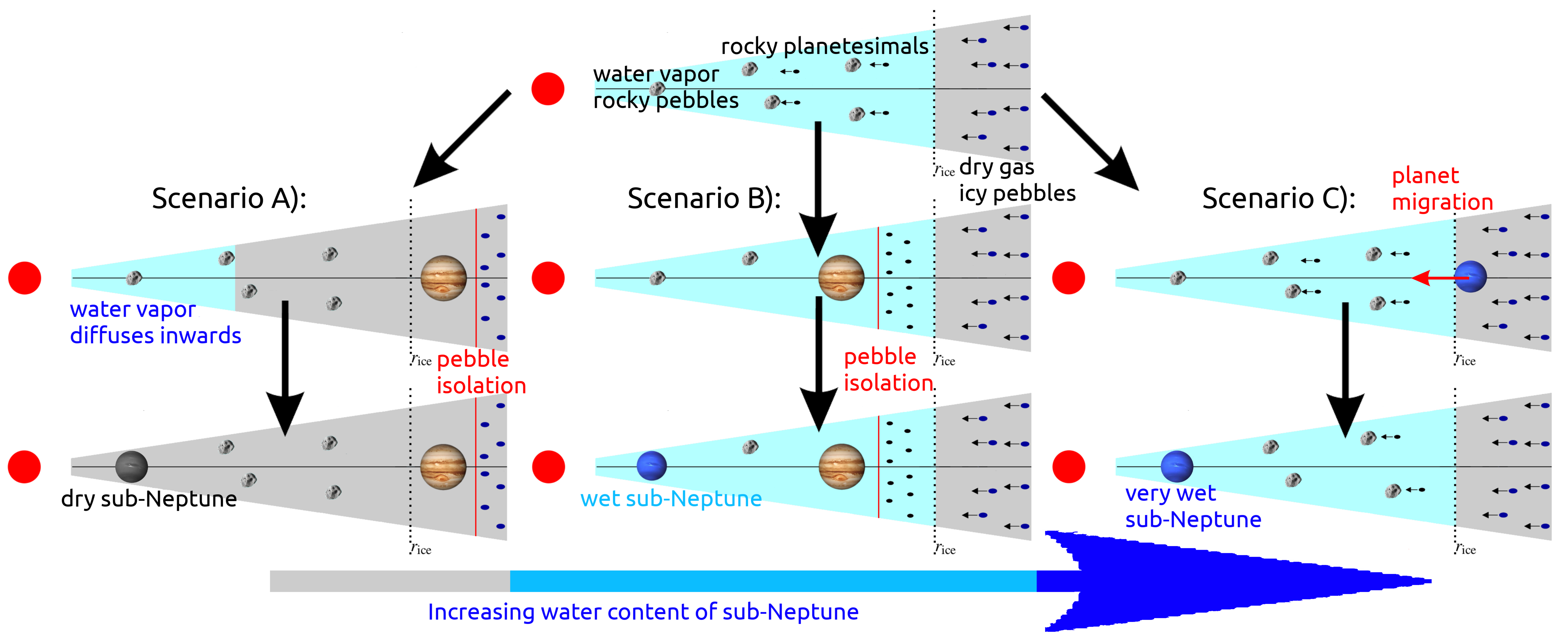} 
 \caption{Cartoon depicting the formation scenarios of the inner sub-Neptunes and their outer giant planets. Pebbles drift inward and release their volatiles at the water ice line, resulting in water vapor in the inner disk, where rocky pebbles are initially available to form rocky planetesimals. In scenario A, the giant planet forms outside the water ice line, blocking water-rich pebbles and thus preventing them from crossing the water ice line. The originally present water vapor diffuses inward and is lost over time. A dry sub-Neptune can then form in the inner disk. In scenario B the giant planet forms inside the water ice line, and thus the pebbles release their water vapor into the disk as they cross the water ice line. This water vapor can then diffuse inward because gas accretion onto giant planets is not 100\% efficient. As a result, planets forming in the inner region can accrete water-rich vapor from the disk and form wet sub-Neptunes. In scenario C the sub-Neptune forms outside the water ice line, where it accretes a large amount of water ice and then migrates toward the inner edge of the disk, resulting in a very wet sub-Neptune. Here, the formation time and location of the giant planet do not matter. A sub-Neptune formed via scenario C could thus contain up to 50\% of its total mass in water ice, while a sub-Neptune formed via scenario B would contain a water-rich atmosphere on top of a rocky core; as such, its overall water mass might be just a few percent. For the sake of simplicity, we do not show the movement of the water ice line over time or water-rich planetesimals. (For a discussion about diffusion, see Appendix~\ref{ap:water}.)
   \label{fig:Cartoon}
   }
\end{figure*}

Here, we discuss how different water contents of close-in sub-Neptunes can arise and what it implies for the formation of these systems. In particular, we focus on three formation scenarios, illustrated in Fig.~\ref{fig:Cartoon}, which can be distinguished by the water content of the inner sub-Neptune. Our scenarios differ in terms of the relative timing of events.  

In the first scenario, the water ice line has been "fossilized" by the rapid growth of Jupiter's core. \citet{2016Icar..267..368M} suggested that the pressure bump exerted by Jupiter's growing core\footnote{In the pebble accretion scenario, this is referred to as the pebble isolation mass \citep{2014A&A...572A..35L, 2018arXiv180102341B, 2018A&A...615A.110A}.} blocks the inward flow of icy pebbles and thus keeps the solid icy material away from the inner system. The water vapor, originally inside Jupiter's orbit, diffuses inward faster than the water ice line moves inward \citep{2006ApJ...640.1115L, 2011ApJ...738..141O, 2015A&A...575A..28B, 2015arXiv150303352B, 2020arXiv200514097S}, preventing recondensation of the water vapor onto interior planetesimals and thus keeping the inner disk dry. In addition, the disk outside Jupiter's core is cold enough that water ice does not evaporate, preventing the diffusion of water vapor into the inner system. This results in a dry region inside Jupiter's orbit. The water ice line is thus fossilized at the original distance where Jupiter first blocked the inward drifting pebbles, and a dry sub-Neptune can form in the inner disk via the accretion of planetesimals (scenario A in Fig.~\ref{fig:Cartoon}). 

In this scenario, it is crucial that the giant planet forms early compared to the inner sub-Neptune, so that the inner disk is dry during the formation of the sub-Neptune. In the Solar System, Jupiter's formation is invoked to explain the differences between carbonaceous and non-carbonaceous chondrites \citep{2017LPI....48.1386K}. In particular, this scenario involves an early formation of Jupiter, which blocks inward drifting pebbles within the first megayear. 

For scenario A, it is thus crucial that the giant planet is able to efficiently block the vast majority of water-ice-rich particles to prevent a contamination of the inner disk with water vapor. The exact efficiency of pebble filtering depends on the planetary mass as well as on the strength of the diffusion (e.g., \citealt{2012ApJ...755....6Z, 2018arXiv180102341B, 2018A&A...615A.110A, 2019ApJ...885...91D}). Two-dimensional simulations of an embedded Jupiter mass planet in a disk with $\alpha=0.001$ indicate that small dust grains could indeed diffuse through the pressure bump and then re-coagulate \citep{2019ApJ...885...91D}; however, from a Solar System perspective, the amount of material that is allowed to diffuse into the inner system should be minimal \citep{2018ApJ...854..153W, 2019AJ....158...55H}. Furthermore, observations of transition disks indicate that grain growth to millimeter size and beyond in the inner disk regions becomes inefficient once particles are blocked at a pressure trap (presumably caused by planets), implying that only a minimal amount of dust could diffuse into the inner disk (e.g., \citealt[][see also Appendix~\ref{ap:water}]{2020ApJ...892..111F, 2021MNRAS.502.5779N}). Scenario A thus represents an idealization, where clearly the observation of a dry (water-poor) sub-Neptune would place very interesting constraints on planet formation theories.

On the other hand, if the giant planet is located inside the water ice line, and thus generates a pressure bump inside the water ice line, inward drifting pebbles can evaporate and enrich the gas with water vapor. The water vapor can then diffuse inward all the way to the central star because gas accretion by giant planets is not 100\% efficient \citep{2006ApJ...641..526L, 2017A&A...598A..80D, 2020A&A...643A.133B}. The forming inner sub-Neptune can accrete water-rich vapor into its atmosphere (scenario B in Fig.~\ref{fig:Cartoon}), resulting in a wet sub-Neptune with an atmosphere composed of up to a few percent water vapor on top of a rocky core.

If the sub-Neptune forms outside the water ice line, it can accrete large amounts of water-rich pebbles and planetesimals \citep{2017A&A...604A...1O, 2019A&A...624A.109B, 2019A&A...627A.149S, 2019A&A...632A...7L, 2020A&A...643L...1V} before it migrates to the inner edge of the protoplanetary disk. As a result, a very wet sub-Neptune is formed (scenario C in Fig.~\ref{fig:Cartoon}), which can consist of up to 50\% water ice. This scenario is independent of the formation location and time of the giant planet, which should form after the inward migration of the sub-Neptune.

Of course, protoplanetary disks evolve over time. They cool due to the reduced accretion rates, which reduce the viscous heating and thus the disk's temperature \citep{2006ApJ...640.1115L, 2011ApJ...738..141O, 2015A&A...575A..28B, 2015arXiv150303352B, 2020arXiv200514097S}. In fact, the water ice line moves within 1 AU in the first megayear in most models, suggesting that inner planets have accreted a large fraction of water. Early giant planet formation, on the other hand, could keep the inner disk dry, even when the disk cools significantly \citep{2016Icar..267..368M}. The water content of inner sub-Neptunes with outer giant planets could thus help to constrain the formation sequence of the planets within the system (Fig.~\ref{fig:Cartoon}).

In Table~\ref{tab:SE} we list all the known super-Earths ($R_{\rm P} < 2 R_{\rm E}$) and sub-Neptunes ($R_{\rm P} > 2 R_{\rm E}$) with outer gas giants, and we list the gas giants in the same system in Table~\ref{tab:CJ}. In total, our list includes 14 systems with a total of 19 super-Earths and sub-Neptunes as well as 19 giant planets. The smallest giant planets listed in Table~\ref{tab:CJ} are around 50 Earth masses (in the 55-Cnc system), which are large enough to reach the pebble isolation mass in the inner regions of the disk \citep{2014A&A...572A..35L, 2018arXiv180102341B, 2018A&A...615A.110A}.

\begin{figure*}
 \centering
 \includegraphics[scale=0.7]{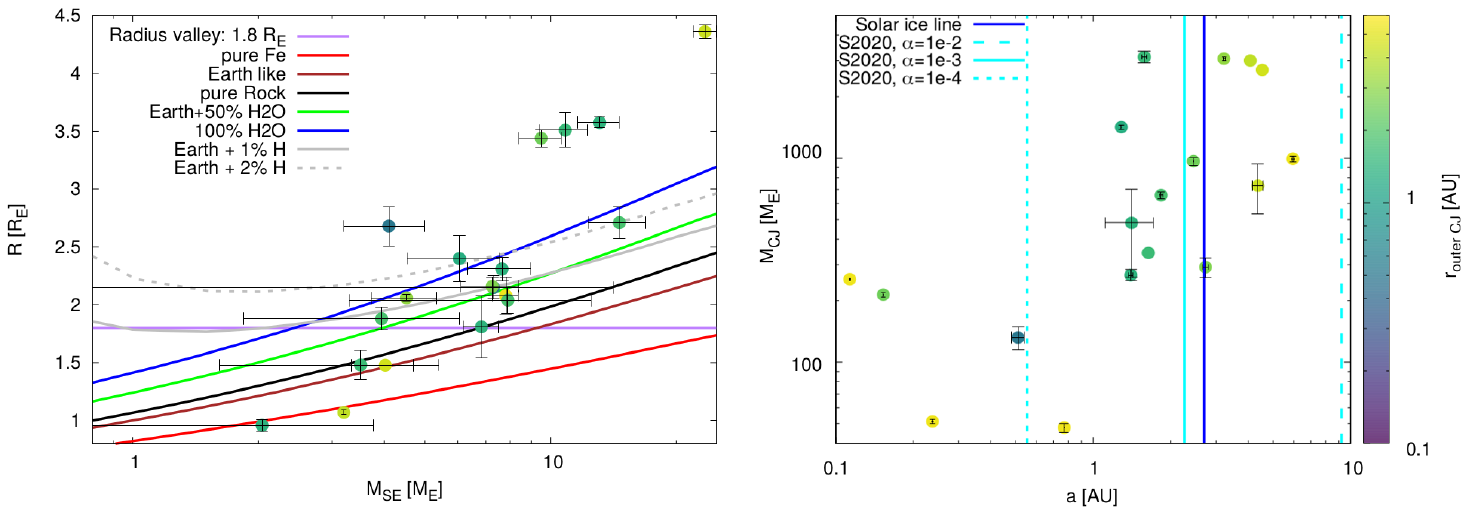}  
 \caption{Observed systems with transit and RV data that contain inner super-Earths and sub-Neptunes and outer gas giants. {\bf Left:} Mass-radius diagram for super-Earths and sub-Neptunes with outer gas giants, color coded by the orbital position of the outermost gas giant in the same system, $r_{\rm outer}$. The mass-radius relation determined by models with fixed composition following the model of \citet{2019PNAS..116.9723Z} is shown.\ Also shown, in purple, is the position of the radius valley  \citep{2017AJ....154..109F}, below which planets are presumably of a rocky nature. {\bf Right:} Orbital distance and masses of the cold Jupiters of the corresponding inner systems shown in the left panel. The planets are color coded by the orbital position of the outermost giant planet. In addition, we show the present-day position of the water ice line in the Solar System (dark blue) and the ice line positions in protoplanetary disk models following \citet{2020arXiv200514097S} with different viscosities.  The two giant planets with masses below 100 Earth masses are 55-Cnc-c and 55-Cnc-f. Some planets do not contain error bars in the database (see Table~\ref{tab:CJ}), and thus have no error bars here.
   \label{fig:MRCJ}
   }
\end{figure*}

Figure~\ref{fig:MRCJ} shows the masses and orbital distances of the massive planets (right) and the mass-radius relations of inner super-Earths and sub-Neptunes (left) from Table.~\ref{tab:SE} compared with interior structure models from \citet{2019PNAS..116.9723Z}. From the planetary mass and radius alone, it is difficult to determine the planetary composition because there are many more planet-building materials than constraints, which results in a degenerate problem (e.g., \citealt{2007Icar..191..453S}). Nevertheless, a few of the planets clearly correspond to a rocky composition. We thus do not expect these planets to harbor any significant amount of water. Recent studies of atmospheric evaporation also predict that small super-Earths ($R_{\rm P} < 2 R_{\rm E}$) have a predominantly rocky nature (e.g., \citealt{2017ApJ...847...29O}), which has been confirmed by observations \citep{2017AJ....154..109F}; these observations have found a valley in the radius distribution of these planets at around $1.8$ Earth radii. However, the composition of larger planets is still unclear and needs to be constrained in detail in the future.

\section{Position of the water ice line}
\label{sec:water}

The position of the water ice line in a protoplanetary disk can be determined by calculating the balance between heating and cooling \citep{2006ApJ...640.1115L, 2011ApJ...738..141O, 2015A&A...575A..28B, 2015arXiv150303352B, 2020arXiv200514097S}. The cooling of the disk depends crucially on the grain opacity, which itself is a function of the grain size. Larger grains have a lower opacity compared to smaller grains, emphasizing the importance of calculating the exact grain size distribution in protoplanetary disks for the thermal structure of the disk.

The grain sizes in protoplanetary disks are determined by an equilibrium between the coagulation and fragmentation of dust grains (e.g., \citealt{2008A&A...480..859B}). The maximal size particles can thus grow to depends on the relative velocities between the grains. The maximal velocities the grains can reach before they fragment is the called ``fragmentation velocity'' and is typically on the order of 1-10 m/s \citep{2015ApJ...798...34G, 2019ApJ...873...58M}.

In \citet{2020arXiv200514097S}, the disk temperature was calculated in 2D radiation hydrodynamical simulations that included a full grain size distribution with a constant fragmentation velocity\footnote{Larger grain fragmentation velocities would lead to larger grain sizes and thus even colder disks (see \citealt{2021arXiv210311995S}).} of 1 m/s. They gave a fit of the water ice line position around a solar mass star as a function of the disk's gas surface density, $\Sigma_{\rm g,0}$, at 1 AU, the $\alpha$-viscosity parameter \citep{1973A&A....24..337S}, and the dust-to-gas ratio, $f_{\rm DG}$, as
\begin{equation}
 \label{eq:ice}
 r_{\rm ice} = 9.2 \left(\frac{\alpha}{0.01}\right)^{0.61} \left(\frac{\Sigma_{\rm g,0}}{1000 {\rm g/cm^2}}\right)^{0.8} \left(\frac{f_{\rm DG}}{0.01}\right)^{0.37} {\rm AU} \ .
\end{equation}
The dependence of the water ice line position on these parameters is explained in the following way: (i) A lower $\alpha$ viscosity results in less turbulence, which then generates less viscous heating, resulting in a colder disk\footnote{A lower turbulence also allows grains to grow larger, reducing the opacity and thus increasing the cooling rates, in turn resulting in an even colder disk.}, (ii) a reduced gas surface density results in less viscous heating and an increased cooling rate, resulting in a cooler disk, and (iii) a higher dust-to-gas ratio increases the opacity, thus decreasing the cooling rate and resulting in a hotter disk. The opacity in this disk model is calculated self-consistently from the full grain size distribution at each radial position.

While Eq.~\ref{eq:ice} is derived for a solar mass star, planetary systems are also found around stars with different masses (e.g., Table~\ref{tab:SE}). A change in the stellar mass automatically implies a change in the stellar luminosity and with this a change in the stellar heating of the disk. However, the transition of the water ice line is within the viscously heated part of the disk \citep{2015A&A...575A..28B, 2020arXiv200514097S, 2021arXiv210311995S}, so a change in the stellar luminosity might only marginally influence the ice line position. Furthermore, the quantitative statements in this section are unaffected by this, and each planetary system would require a detailed analysis once the inner sub-Neptune is characterized.

We plot in Fig.~\ref{fig:MRCJ} the giant planets of the same systems as well as the position of the water ice lines for different levels of viscosity ($\alpha= 10^{-4}$, $10^{-3}$, and $10^{-2}$), a disk gas surface density of 1000g/cm$^2$, and a dust-to-gas ratio of 0.01 following Eq.~\ref{eq:ice}. The exact dust-to-gas ratio in the disk is a function of the host star metallicity [Fe/H] (see Table~\ref{tab:CJ}), but we do not take these effects into account here for the sake of simplicity and because the position of the ice line depends more strongly on the disk's viscosity and gas surface density. Clearly, a reduction in the disk's viscosity results in a water ice line closer to the host star. The level of the disk turbulence remains a subject of many studies, but observational evidence seems to point to turbulence levels corresponding to $\alpha < 10^{-3}$ \citep{2018ApJ...856..117F, 2018ApJ...869L..46D}. In addition, magnetohydrodynamic  simulations point to low levels of turbulence in the disk \citep{2016ApJ...821...80B, 2016arXiv160900437S, 2017arXiv170700729B} as well.

As the disk evolves, the disk's accretion rate lowers over time \citep{1998ApJ...495..385H}, and with it the gas surface density.\ This results in a cooling of the disk over time, which causes the aforementioned inward motion of the water ice line \citep{2011ApJ...738..141O, 2015A&A...575A..28B}. In order for the fossilization of the water ice line to take place, the giant planet must remain outside the water ice line at all times (see scenario A in Fig.~\ref{fig:Cartoon}). Only then is it assured that all the inward drifting icy pebbles can be blocked by the giant planet. (If the water ice line were outside the giant planet, water-rich pebbles drifting inward would evaporate before being blocked by the giant planet.) In this scenario, water-rich gas can then diffuse inward, and some of the gas will pass the giant planet because gas accretion is not 100\% efficient (e.g., \citealt{2006ApJ...641..526L, 2017A&A...598A..80D}); this gas could then be accreted by the inner planets, resulting in sub-Neptunes with water-rich atmospheres (scenario B in Fig.~\ref{fig:Cartoon}).

\section{Planet migration and the timing of planet formation}
\label{sec:migration}

Planets above $\approx$0.1 Earth masses start to migrate through the disk (for a review, see \citealt{2013arXiv1312.4293B}). Planets of a few Earth masses migrate in the type-I regime, in which such planets can easily migrate all the way to the inner edge of the disk and form a chain of super-Earth planets \citep{2007ApJ...654.1110T, 2010ApJ...719..810I, 2014arXiv1407.6011C, 2016MNRAS.457.2480C, 2017MNRAS.470.1750I, 2018arXiv181001389O, 2019arXiv190208772I}. These super-Earths, if they form outside the water ice line, could actually harbor a large water fraction \citep{2019A&A...624A.109B, 2019arXiv190208772I, 2019A&A...627A.149S, 2020A&A...643L...1V}. The exact water fraction of the sub-Neptunes therefore depends on a complex interplay between the planetary growth rate, planetary migration, and the evolution of the water ice line over time \citep{2019A&A...624A.109B}, as well as on the scattering and merging of planets \citep{2018MNRAS.479L..81R}.

Planets that start to accrete gas efficiently start to open gaps in the protoplanetary disk and then migrate on a much longer timescale that depends on the depth of the gap \citep{2018arXiv180511101K}. More massive planets open deeper gaps and thus migrate more slowly, also influenced by the gas accretion rate itself \citep{2017Icar..285..145C, 2020A&A...643A.133B, 2020MNRAS.tmp.3444N}. The migration speed in the type-II regime, and thus also the distance over which planets migrate, also crucially depends on the disk's viscosity. In the case of high viscosity ($\alpha >0.005$), giant planets can migrate over more than 20 AU, until the disk dissipates \citep{2015A&A...582A.112B}. In disks with low viscosity (e.g., $\alpha=10^{-4}$), giant planets could still migrate over a distance of a few AU during their formation (e.g., \citealt{2019A&A...623A..88B}).

As migration is a universal process and is unavoidable, the giant planets probably formed farther away from their host star than their current position. However, what matters is the position of the planet relative to the position of the water ice line. As long as the giant planet is always outside the water ice line, the water ice line will be fossilized as soon as the planet becomes massive enough to open a gap in the pebble disk \citep{2006A&A...453.1129P}. This can happen at the already low masses of around 20 Earth masses \citep{2014A&A...572A..35L, 2018arXiv180102341B}. 

However, if a planet outside the water ice line grows very slowly, it will block the pebble flux too late, allowing a large amount of water-rich pebbles to drift inward. These pebbles can then release water into the gas phase once they cross the water ice line. This could lead to sub-Neptunes formed in the inner disk that accrete water vapor from the disk. Alternatively, if planetary cores grow too slowly at the water ice line, they are subject to type-I migration for a long time and might migrate into the inner disk before they become gas giants, forming very water-rich sub-Neptunes \citep[][see scenario C in Fig.~\ref{fig:Cartoon}]{2019arXiv190208772I, 2019A&A...623A..88B, 2020A&A...643L...1V}. The fast growth of the giant planet is thus key in keeping the inner system dry and thus allowing the formation of a dry sub-Neptune. 

Planetary growth is enhanced in systems with super-solar metallicity because more building blocks are available \citep{2015A&A...582A.112B, 2018MNRAS.474..886N, 2012A&A...541A..97M}. In Table~\ref{tab:CJ} we also list the host star metallicity, which clearly indicates that most of the systems have a super-solar metallicity; this in turn indicates that the fast growth of giant planets is possible, in line with the giant-planet host star metallicity correlation \citep{2004A&A...415.1153S, 2005ApJ...622.1102F, J2010}. 

Additionally, growing giant planets close to the ice line can scatter water-rich planetesimals into the inner system, where they can be accreted by inner planets. However, these contributions to the water content are generally small, allowing them to explain the low water content of the Earth \citep{2017Icar..297..134R}.

On the other hand, if the giant planet is always inside the water ice line, then the fossilization of the water ice line cannot take place and water-rich gas could pass the gas giant, allowing the accretion of water vapor onto sub-Neptunes in the inner region (scenario B in Fig.~\ref{fig:Cartoon}). Alternatively, sub-Neptunes could form outside the gas giant, allowing a water-rich composition. However, the giant planet acts as a barrier to the inward migrating small planets \citep{2015ApJ...800L..22I}, preventing the water-rich sub-Neptunes from migrating from the inner disks in great numbers.

In Appendix~\ref{ap:water} we show the water content in the gas phase for our three scenarios from a model that includes pebble growth and drift as well as evaporation and condensation at ice lines. Our model shows that the water content of the inner disk can be increased due to pebble evaporation (see also \citealt{2017MNRAS.469.3994B, 2020ApJ...903..124B, 2021arXiv210301793M}). Sub-Neptunes that accrete their gaseous envelope in the inner disk could thus only have a water content of up to a few percent (scenario B); this clearly distinguishes them from sub-Neptunes that form beyond the water ice line and then migrate inward (scenario C), which should have a much larger water content of several tens of percent \citep{2019A&A...624A.109B, 2019arXiv190208772I, 2020A&A...643L...1V}, depending on the original water content of the system \citep{2020A&A...633A..10B}.

\section{Rocky or water world}
\label{sec:rocky}

In the case a system with an outer giant planet and an inner, dry (or wet) sub-Neptune, we can constrain the formation history of the system within our model. We discuss our results under the assumption that the gas surface density at the time the giant planets formed was 1000g/cm$^2$ at 1 AU. This is already below the gas surface density of the Minimum Mass Solar Nebula \citep{1977Ap&SS..51..153W, 1981PThPS..70...35H}; however, giant planets take at least a few hundred kiloyears to form, and as such the disk is already evolved\footnote{A larger gas surface density at the time when the giant planet blocks the pebble flux shifts the ice line farther out, but in the case of low viscosity ($\alpha=10^{-4}$) a very large gas surface density of 6000 g/cm$^2$ is needed to reach the same ice line position as for $\alpha=10^{-3}$ with 1000g/cm$^2$.}.

The position of the giant planet determines if water-rich material can make it into the inner disk (Fig.~\ref{fig:Cartoon}). If the sub-Neptune forms beyond the water ice line (scenario C in Fig.~\ref{fig:Cartoon}), it will be very water rich and the position of the giant planet relative to the ice line does not matter for the water content of the sub-Neptune.

The following planets are clearly rock dominated because their mass-radius relationships fall below the pure-rocky composition (Fig.~\ref{fig:MRCJ}): WASP-47e, K-68c, K-407b, K-97b, K-93b, and 55-Cnc-e. These planets should thus harbor only minimal amounts of water.\  For 55-Cnc-eas, this was shown with several observation campaigns  \citep{2017AJ....153..268E, 2020arXiv200703115J} and was concluded from the simulations of \citet{2019MNRAS.484..712D}. For the sake of simplicity, we thus exclude these planets from further discussion.

The WASP-47 system presents a very special case because it harbors two sub-Neptunes and two giant planets. In particular, the giant planet WASP-47b orbits its host star on only a 4 day period, while the other giant planet (WASP-47c) orbits the star in nearly 600 days (distance of 1.41 AU). If WASP-47c had always been outside the water ice line and formed very fast and early, then the inner planets should not harbor any water. This is the case for WASP-47e, which shows a purely rocky composition (Fig.~\ref{fig:MRCJ}), but it is unclear for WASP-47b and WASP-47d. WASP-47b is a hot Jupiter and could have accreted water vapor into its atmosphere if the outer giant planet formed late; this makes WASP-47b a very interesting target for atmospheric characterization.

The K-289 system also features a giant planet very close to the star (at 0.51 AU), which may have always been inside the water ice line; if so, it would have been unable to fossilize the water ice line, allowing water vapor to diffuse into the inner disk. The two inner planets have no conclusive mass-radius relations (Fig.~\ref{fig:MRCJ}), but according to our theory these planets should contain water.

The other planetary systems (Tables~\ref{tab:SE} and \ref{tab:CJ}) contain gas giants on orbital distances of at least 1-2 AU; for these systems it is unclear if the giant planets have always been inside or outside the water ice line (since the water ice line evolves; not shown in Fig.~\ref{fig:MRCJ}). Under the assumption that the giant planets did not migrate outward and grew fast and early in the disk, we list the predictions of our model for the water content of the inner planets in Table~\ref{tab:SEcomp} as a function of the $\alpha$ viscosity parameter, which determines the position of the water ice line.

Observations of HAT-P-11b seem to indicate water vapor inside the planetary atmosphere \citep{2014Natur.513..526F, 2019AJ....158..244C}; however, the water content seems to be quite low. Within our model, this would favor formation scenario B, implying an ice line initially far away from the host star. Observations of $\pi$-Men-c by \citet{2021ApJ...907L..36M} indicate a thick atmosphere with abundant volatiles, indicating that $\pi$-Men-c may have formed via our formation scenario C. Clearly, better constraints for these targets are needed to give clear constraints to our planet formation scenarios.

\begin{table}
\centering
\begin{tabular}{c|c|c|c}
\hline
Name    &       $\alpha$=0.01 & $\alpha$=0.001& $\alpha=10^{-4}$ \\ \hline \hline
K-68b   &       wet&    wet&    {\bf dry} \\
K-48b   &       wet&    wet&    {\bf dry}  \\
K-48c   &       wet&    wet&    {\bf dry}  \\
K-48d   &       wet &   wet&    {\bf dry}  \\
K-94b   &       wet&    wet&    {\bf dry}  \\
$\pi$-Men-c&    wet&    {\bf dry}&      {\bf dry} \\    
WASP-47d&       wet&    wet&    {\bf dry} \\
K-454b  &       wet&    wet&    {\bf dry} \\
K-289b  &       wet&    wet&    wet \\
K-289c  &       wet&    wet&    wet \\
HD-86226c &   wet& {\bf dry}&   {\bf dry} \\
HAT-P-11b & wet& {\bf dry}&     {\bf dry} \\
K-88b   &   wet& {\bf dry}&   {\bf dry} \\
\hline
\end{tabular}
\caption[Super-Earths]{Predictions of the water content of sub-Neptunes from our model if the system formed via scenario A or scenario B (Fig.~\ref{fig:Cartoon}) and depending on the $\alpha$ value for the ice line position (always using $\Sigma_{\rm g}$ = 1000g/cm$^2$). We only list the planets that have a mean density lower than pure rock (Fig.~\ref{fig:MRCJ}). The exact water content of the inner planet then depends on if the sub-Neptune is formed via scenario B or C (see Appendix~\ref{ap:water}).}
\label{tab:SEcomp}
\end{table}

Our model predicts that the water content of inner sub-Neptunes decreases if the water ice line is closer to the central star (low $\alpha$). Our model only predicts a few systems where the inner planets should be water poor, assuming that moderate $\alpha$ values determine the position of the water ice line. Only for very low $\alpha$ values, and thus negligible viscous heating, does our model predict a majority of water-poor inner planets. The detection of water-poor inner sub-Neptunes would thus place the strongest constraint on planet formation models.

The exact water content of an inner sub-Neptune planet can thus tell us a great deal about the formation speed and position (relative to the water ice line) of its outer giant planet, constraining planet formation theories. Additionally, the atmospheric C/O ratio of sub-Neptunes could help to constrain the formation location and timing of the giant planets even further because carbon-bearing species can also be blocked by growing giant planets. This, of course, mostly applies to CO$_2$ rather than CH$_4$ or CO because of the closer proximity of the CO$_2$ ice line (at around 70K) to the host star and to the final giant planet position. Future observations of the water content of inner sub-Neptune planets are thus crucial for constraining planet formation theories.

\begin{acknowledgements}

B.B., thanks the European Research Council (ERC Starting Grant 757448-PAMDORA) for their financial support. S.N.R. is grateful to the CNRS's PNP program for support. A.D.S is part of the CHAMELEON MC ITN EJD which received funding from the European Union’s Horizon 2020 research and innovation programme under the Marie Sklodowska-Curie grant agreement no. 860470.

\end{acknowledgements}

\bibliographystyle{aa}
\bibliography{Stellar}

\begin{thebibliography}{81}
\expandafter\ifx\csname natexlab\endcsname\relax\def\natexlab#1{#1}\fi

\bibitem[{{Ataiee} {et~al.}(2018){Ataiee}, {Baruteau}, {Alibert}, \&
  {Benz}}]{2018A&A...615A.110A}
{Ataiee}, S., {Baruteau}, C., {Alibert}, Y., \& {Benz}, W. 2018, \aap, 615,
  A110

\bibitem[{Bai(2016)}]{2016ApJ...821...80B}
Bai, X.~N. 2016, ApJ, 821, id.80

\bibitem[{Bai(2017)}]{2017arXiv170700729B}
Bai, X.~N. 2017, ApJ, 845, id.75

\bibitem[{Bailli{\'e} {et~al.}(2015)Bailli{\'e}, Charnoz, \&
  Pantin}]{2015arXiv150303352B}
Bailli{\'e}, K., Charnoz, S., \& Pantin, {\'E}. 2015, astro-ph.EP

\bibitem[{{Banzatti} {et~al.}(2020){Banzatti}, {Pascucci}, {Bosman}, {Pinilla},
  {Salyk}, {Herczeg}, {Pontoppidan}, {Vazquez}, {Watkins}, {Krijt}, {Hendler},
  \& {Long}}]{2020ApJ...903..124B}
{Banzatti}, A., {Pascucci}, I., {Bosman}, A.~D., {et~al.} 2020, \apj, 903, 124

\bibitem[{{Baruteau} {et~al.}(2014){Baruteau}, {Crida}, {Paardekooper},
  {Masset}, {Guilet}, {Bitsch}, Nelson, {Kley}, \&
  {Papaloizou}}]{2013arXiv1312.4293B}
{Baruteau}, C., {Crida}, A., {Paardekooper}, S.~J., {et~al.} 2014, in
  {Protostars and Planets VI}, arXiv:1312.4293

\bibitem[{{Batalha} {et~al.}(2019){Batalha}, {Lewis}, {Fortney}, {Batalha},
  {Kempton}, {Lewis}, \& {Line}}]{2019ApJ...885L..25B}
{Batalha}, N.~E., {Lewis}, T., {Fortney}, J.~J., {et~al.} 2019, \apjl, 885, L25

\bibitem[{{Benneke} {et~al.}(2019){Benneke}, {Wong}, {Piaulet}, {Knutson},
  {Lothringer}, {Morley}, {Crossfield}, {Gao}, {Greene}, {Dressing},
  {Dragomir}, {Howard}, {McCullough}, {Kempton}, {Fortney}, \&
  {Fraine}}]{2019ApJ...887L..14B}
{Benneke}, B., {Wong}, I., {Piaulet}, C., {et~al.} 2019, \apjl, 887, L14

\bibitem[{{Bergez-Casalou} {et~al.}(2020){Bergez-Casalou}, {Bitsch}, {Pierens},
  {Crida}, \& {Raymond}}]{2020A&A...643A.133B}
{Bergez-Casalou}, C., {Bitsch}, B., {Pierens}, A., {Crida}, A., \& {Raymond},
  S.~N. 2020, \aap, 643, A133

\bibitem[{Birnstiel {et~al.}(2012)Birnstiel, Klahr, \&
  Ercolano}]{2012A&A...539A.148B}
Birnstiel, T., Klahr, H., \& Ercolano, B. 2012, A\&A, 539, id.A148

\bibitem[{{Bitsch} \& {Battistini}(2020)}]{2020A&A...633A..10B}
{Bitsch}, B. \& {Battistini}, C. 2020, \aap, 633, A10

\bibitem[{{Bitsch} {et~al.}(2019{\natexlab{a}}){Bitsch}, {Izidoro}, {Johansen},
  {Raymond}, {Morbidelli}, {Lambrechts}, \& {Jacobson}}]{2019A&A...623A..88B}
{Bitsch}, B., {Izidoro}, A., {Johansen}, A., {et~al.} 2019{\natexlab{a}}, \aap,
  623, A88

\bibitem[{Bitsch {et~al.}(2015{\natexlab{a}})Bitsch, Johansen, Lambrechts, \&
  Morbidelli}]{2015A&A...575A..28B}
Bitsch, B., Johansen, A., Lambrechts, M., \& Morbidelli, A. 2015{\natexlab{a}},
  A\&A, 575, id.A28

\bibitem[{Bitsch {et~al.}(2015{\natexlab{b}})Bitsch, Lambrechts, \&
  Johansen}]{2015A&A...582A.112B}
Bitsch, B., Lambrechts, M., \& Johansen, A. 2015{\natexlab{b}}, A\&A, 582,
  id.A112

\bibitem[{{Bitsch} {et~al.}(2018){Bitsch}, {Morbidelli}, {Johansen}, {Lega},
  {Lambrechts}, \& {Crida}}]{2018arXiv180102341B}
{Bitsch}, B., {Morbidelli}, A., {Johansen}, A., {et~al.} 2018, A\&A, 612,
  id.A30

\bibitem[{{Bitsch} {et~al.}(2019{\natexlab{b}}){Bitsch}, {Raymond}, \&
  {Izidoro}}]{2019A&A...624A.109B}
{Bitsch}, B., {Raymond}, S.~N., \& {Izidoro}, A. 2019{\natexlab{b}}, \aap, 624,
  A109

\bibitem[{Booth {et~al.}(2017)Booth, Clarke, Madhusudhan, \&
  Ilee}]{2017MNRAS.469.3994B}
Booth, R.~A., Clarke, C.~J., Madhusudhan, N., \& Ilee, J.~D. 2017, MNRAS, 469,
  p.3994

\bibitem[{Brauer {et~al.}(2008)Brauer, Dullemond, \&
  Henning}]{2008A&A...480..859B}
Brauer, F., Dullemond, C., \& Henning, T. 2008, A\&A, 480, pp.859

\bibitem[{{Chachan} {et~al.}(2019){Chachan}, {Knutson}, {Gao}, {Kataria},
  {Wong}, {Henry}, {Benneke}, {Zhang}, {Barstow}, {Bean}, {Mikal-Evans},
  {Lewis}, {Mansfield}, {L{\'o}pez-Morales}, {Nikolov}, {Sing}, \&
  {Wakeford}}]{2019AJ....158..244C}
{Chachan}, Y., {Knutson}, H.~A., {Gao}, P., {et~al.} 2019, \aj, 158, 244

\bibitem[{{Coleman} \& {Nelson}(2016)}]{2016MNRAS.457.2480C}
{Coleman}, G. A.~L. \& {Nelson}, R.~P. 2016, \mnras, 457, 2480

\bibitem[{Cossou {et~al.}(2014)Cossou, Raymond, Hersant, \&
  Pierens}]{2014arXiv1407.6011C}
Cossou, C., Raymond, S.~N., Hersant, F., \& Pierens, A. 2014, A\&A, 569, id.
  A56

\bibitem[{Crida \& Bitsch(2017)}]{2017Icar..285..145C}
Crida, A. \& Bitsch, B. 2017, Icarus, 285, p. 145

\bibitem[{{Crida} {et~al.}(2006){Crida}, {Morbidelli}, \&
  {Masset}}]{2006Icar..181..587C}
{Crida}, A., {Morbidelli}, A., \& {Masset}, F. 2006, Icarus, 181, 587

\bibitem[{{Dorn} {et~al.}(2019){Dorn}, {Harrison}, {Bonsor}, \&
  {Hands}}]{2019MNRAS.484..712D}
{Dorn}, C., {Harrison}, J.~H.~D., {Bonsor}, A., \& {Hands}, T.~O. 2019, \mnras,
  484, 712

\bibitem[{{Dr{\k{a}}{\.z}kowska} {et~al.}(2019){Dr{\k{a}}{\.z}kowska}, {Li},
  {Birnstiel}, {Stammler}, \& {Li}}]{2019ApJ...885...91D}
{Dr{\k{a}}{\.z}kowska}, J., {Li}, S., {Birnstiel}, T., {Stammler}, S.~M., \&
  {Li}, H. 2019, \apj, 885, 91

\bibitem[{{Dullemond} {et~al.}(2018){Dullemond}, {Birnstiel}, {Huang},
  {Kurtovic}, {Andrews}, {Guzm{\'a}n}, {P{\'e}rez}, {Isella}, {Zhu}, {Benisty},
  {Wilner}, {Bai}, {Carpenter}, {Zhang}, \& {Ricci}}]{2018ApJ...869L..46D}
{Dullemond}, C.~P., {Birnstiel}, T., {Huang}, J., {et~al.} 2018, \apjl, 869,
  L46

\bibitem[{{D{\"u}rmann} \& {Kley}(2017)}]{2017A&A...598A..80D}
{D{\"u}rmann}, C. \& {Kley}, W. 2017, \aap, 598, A80

\bibitem[{{Esteves} {et~al.}(2017){Esteves}, {de Mooij}, {Jayawardhana},
  {Watson}, \& {de Kok}}]{2017AJ....153..268E}
{Esteves}, L.~J., {de Mooij}, E. J.~W., {Jayawardhana}, R., {Watson}, C., \&
  {de Kok}, R. 2017, \aj, 153, 268

\bibitem[{Fischer \& Valenti(2005)}]{2005ApJ...622.1102F}
Fischer, D.~A. \& Valenti, J. 2005, ApJ, 622, pp. 1102

\bibitem[{{Flaherty} {et~al.}(2018){Flaherty}, {Hughes}, {Teague}, {Simon},
  {Andrews}, \& {Wilner}}]{2018ApJ...856..117F}
{Flaherty}, K.~M., {Hughes}, A.~M., {Teague}, R., {et~al.} 2018, \apj, 856, 117

\bibitem[{{Fraine} {et~al.}(2014){Fraine}, {Deming}, {Benneke}, {Knutson},
  {Jord{\'a}n}, {Espinoza}, {Madhusudhan}, {Wilkins}, \&
  {Todorov}}]{2014Natur.513..526F}
{Fraine}, J., {Deming}, D., {Benneke}, B., {et~al.} 2014, \nat, 513, 526

\bibitem[{{Francis} \& {van der Marel}(2020)}]{2020ApJ...892..111F}
{Francis}, L. \& {van der Marel}, N. 2020, \apj, 892, 111

\bibitem[{{Fulton} {et~al.}(2017){Fulton}, {Petigura}, {Howard}, {Isaacson},
  {Marcy}, {Cargile}, {Hebb}, {Weiss}, {Johnson}, {Morton}, {Sinukoff},
  {Crossfield}, \& {Hirsch}}]{2017AJ....154..109F}
{Fulton}, B.~J., {Petigura}, E.~A., {Howard}, A.~W., {et~al.} 2017, \aj, 154,
  109

\bibitem[{Gundlach \& Blum(2015)}]{2015ApJ...798...34G}
Gundlach, B. \& Blum, J. 2015, ApJ, 798, id. 34

\bibitem[{{Hartmann} {et~al.}(1998){Hartmann}, {Calvet}, {Gullbring}, \&
  {D'Alessio}}]{1998ApJ...495..385H}
{Hartmann}, L., {Calvet}, N., {Gullbring}, E., \& {D'Alessio}, P. 1998, ApJ,
  495, p.385

\bibitem[{{Haugb{\o}lle} {et~al.}(2019){Haugb{\o}lle}, {Weber}, {Wielandt},
  {Ben\'{\i}tez-Llambay}, {Bizzarro}, {Gressel}, \&
  {Pessah}}]{2019AJ....158...55H}
{Haugb{\o}lle}, T., {Weber}, P., {Wielandt}, D.~P., {et~al.} 2019, \aj, 158, 55

\bibitem[{{Hayashi}(1981)}]{1981PThPS..70...35H}
{Hayashi}, C. 1981, Progress of Theoretical Physics Supplement, 70, pp.35

\bibitem[{{Ida} \& {Lin}(2010)}]{2010ApJ...719..810I}
{Ida}, S. \& {Lin}, D.~N.~C. 2010, \apj, 719, 810

\bibitem[{{Izidoro} {et~al.}(2019){Izidoro}, {Bitsch}, {Raymond}, {Johansen},
  {Morbidelli}, {Lambrechts}, \& {Jacobson}}]{2019arXiv190208772I}
{Izidoro}, A., {Bitsch}, B., {Raymond}, S.~N., {et~al.} 2019, arXiv e-prints

\bibitem[{Izidoro {et~al.}(2017)Izidoro, Ogihara, Raymond, Morbidelli, Pierens,
  Bitsch, Cossou, \& Hersant}]{2017MNRAS.470.1750I}
Izidoro, A., Ogihara, M., Raymond, S.~N., {et~al.} 2017, MNRAS, 470, pp. 1750

\bibitem[{{Izidoro} {et~al.}(2015){Izidoro}, {Raymond}, {Morbidelli},
  {Hersant}, \& {Pierens}}]{2015ApJ...800L..22I}
{Izidoro}, A., {Raymond}, S.~N., {Morbidelli}, A., {Hersant}, F., \& {Pierens},
  A. 2015, \apjl, 800, L22

\bibitem[{{Jindal} {et~al.}(2020){Jindal}, {de Mooij}, {Jayawardhana},
  {Deibert}, {Brogi}, {Rustamkulov}, {Fortney}, {Hood}, \&
  {Morley}}]{2020arXiv200703115J}
{Jindal}, A., {de Mooij}, E.~J.~W., {Jayawardhana}, R., {et~al.} 2020, arXiv
  e-prints, arXiv:2007.03115

\bibitem[{{Johnson} {et~al.}(2010){Johnson}, {Aller}, {Howard}, \&
  {Crepp}}]{J2010}
{Johnson}, J.~A., {Aller}, K.~M., {Howard}, A.~W., \& {Crepp}, J.~R. 2010,
  \pasp, 122, 905

\bibitem[{{Kanagawa} {et~al.}(2018){Kanagawa}, {Tanaka}, \&
  {Szuszkiewicz}}]{2018arXiv180511101K}
{Kanagawa}, K.~D., {Tanaka}, H., \& {Szuszkiewicz}, E. 2018, ApJ, 861, id.140

\bibitem[{{Kempton} {et~al.}(2018){Kempton}, {Bean}, {Louie}, {Deming}, {Koll},
  {Mansfield}, {Christiansen}, {L{\'o}pez-Morales}, {Swain}, {Zellem},
  {Ballard}, {Barclay}, {Barstow}, {Batalha}, {Beatty}, {Berta-Thompson},
  {Birkby}, {Buchhave}, {Charbonneau}, {Cowan}, {Crossfield}, {de Val-Borro},
  {Doyon}, {Dragomir}, {Gaidos}, {Heng}, {Hu}, {Kane}, {Kreidberg}, {Mallonn},
  {Morley}, {Narita}, {Nascimbeni}, {Pall{\'e}}, {Quintana}, {Rauscher},
  {Seager}, {Shkolnik}, {Sing}, {Sozzetti}, {Stassun}, {Valenti}, \& {von
  Essen}}]{2018PASP..130k4401K}
{Kempton}, E. M.~R., {Bean}, J.~L., {Louie}, D.~R., {et~al.} 2018, \pasp, 130,
  114401

\bibitem[{{Kreidberg} {et~al.}(2020){Kreidberg}, {Molli{\`e}re}, {Crossfield},
  {Thorngren}, {Kawashima}, {Morley}, {Benneke}, {Mikal-Evans}, {Berardo},
  {Kosiarek}, {Gorjian}, {Ciardi}, {Christiansen}, {Dragomir}, {Dressing},
  {Fortney}, {Fulton}, {Greene}, {Hardegree-Ullman}, {Howard}, {Howell},
  {Isaacson}, {Krick}, {Livingston}, {Lothringer}, {Morales}, {Petigura},
  {Rodriguez}, {Schlieder}, \& {Weiss}}]{2020arXiv200607444K}
{Kreidberg}, L., {Molli{\`e}re}, P., {Crossfield}, I. J.~M., {et~al.} 2020,
  arXiv e-prints, arXiv:2006.07444

\bibitem[{Kruijer {et~al.}(2017)Kruijer, Kleine, Burkhardt, \&
  Budde}]{2017LPI....48.1386K}
Kruijer, T.~S., Kleine, T., Burkhardt, C., \& Budde, G. 2017, LPI, 48th
  Conference, id.1386

\bibitem[{Lambrechts {et~al.}(2014)Lambrechts, Johansen, \&
  Morbidelli}]{2014A&A...572A..35L}
Lambrechts, M., Johansen, A., \& Morbidelli, A. 2014, A\&A, 572, id. A35

\bibitem[{{Lecar} {et~al.}(2006){Lecar}, {Podolak}, {Sasselov}, \&
  {Chiang}}]{2006ApJ...640.1115L}
{Lecar}, M., {Podolak}, M., {Sasselov}, D., \& {Chiang}, E. 2006, \apj, 640,
  1115

\bibitem[{{Liu} {et~al.}(2019){Liu}, {Lambrechts}, {Johansen}, \&
  {Liu}}]{2019A&A...632A...7L}
{Liu}, B., {Lambrechts}, M., {Johansen}, A., \& {Liu}, F. 2019, \aap, 632, A7

\bibitem[{Lubow \& D'Angelo(2006)}]{2006ApJ...641..526L}
Lubow, S.~H. \& D'Angelo, G. 2006, ApJ, 641, pp.526

\bibitem[{Morbidelli {et~al.}(2016)Morbidelli, Bitsch, Crida, Gounelle,
  Guillot, Jacobson, Johansen, Lambrechts, \& Lega}]{2016Icar..267..368M}
Morbidelli, A., Bitsch, B., Crida, A., {et~al.} 2016, Icarus, 267, p. 368

\bibitem[{{Mordasini} {et~al.}(2012){Mordasini}, {Alibert}, {Benz}, {Klahr}, \&
  {Henning}}]{2012A&A...541A..97M}
{Mordasini}, C., {Alibert}, Y., {Benz}, W., {Klahr}, H., \& {Henning}, T. 2012,
  \aap, 541, A97

\bibitem[{{Mousis} {et~al.}(2021){Mousis}, {Aguichine}, {Bouquet}, {Lunine},
  {Danger}, {Mandt}, \& {Luspay-Kuti}}]{2021arXiv210301793M}
{Mousis}, O., {Aguichine}, A., {Bouquet}, A., {et~al.} 2021, arXiv e-prints,
  arXiv:2103.01793

\bibitem[{{Mu{\~n}oz} {et~al.}(2021){Mu{\~n}oz}, {Fossati}, {Youngblood},
  {Nettelmann}, {Gandolfi}, {Cabrera}, \& {Rauer}}]{2021ApJ...907L..36M}
{Mu{\~n}oz}, A.~G., {Fossati}, L., {Youngblood}, A., {et~al.} 2021, \apjl, 907,
  L36

\bibitem[{{Musiolik} \& {Wurm}(2019)}]{2019ApJ...873...58M}
{Musiolik}, G. \& {Wurm}, G. 2019, \apj, 873, 58

\bibitem[{{Ndugu} {et~al.}(2018){Ndugu}, {Bitsch}, \&
  {Jurua}}]{2018MNRAS.474..886N}
{Ndugu}, N., {Bitsch}, B., \& {Jurua}, E. 2018, \mnras, 474, 886

\bibitem[{{Ndugu} {et~al.}(2021){Ndugu}, {Bitsch}, {Morbidelli}, {Crida}, \&
  {Jurua}}]{2021MNRAS.501.2017N}
{Ndugu}, N., {Bitsch}, B., {Morbidelli}, A., {Crida}, A., \& {Jurua}, E. 2021,
  \mnras, 501, 2017

\bibitem[{{Norfolk} {et~al.}(2021){Norfolk}, {Maddison}, {Pinte}, {van der
  Marel}, {Booth}, {Francis}, {Gonzalez}, {M{\'e}nard}, {Wright}, {van der
  Plas}, \& {Garg}}]{2021MNRAS.502.5779N}
{Norfolk}, B.~J., {Maddison}, S.~T., {Pinte}, C., {et~al.} 2021, \mnras, 502,
  5779

\bibitem[{{Ogihara} \& {Hori}(2018)}]{2018arXiv181001389O}
{Ogihara}, M. \& {Hori}, Y. 2018, ApJ, 867, id.127

\bibitem[{{Oka} {et~al.}(2011){Oka}, {Nakamoto}, \&
  {Ida}}]{2011ApJ...738..141O}
{Oka}, A., {Nakamoto}, T., \& {Ida}, S. 2011, ApJ, 738, id.141

\bibitem[{{Ormel} {et~al.}(2017){Ormel}, {Liu}, \&
  {Schoonenberg}}]{2017A&A...604A...1O}
{Ormel}, C.~W., {Liu}, B., \& {Schoonenberg}, D. 2017, \aap, 604, A1

\bibitem[{{Owen} \& {Wu}(2017)}]{2017ApJ...847...29O}
{Owen}, J.~E. \& {Wu}, Y. 2017, \apj, 847, 29

\bibitem[{Paardekooper \& Mellema(2006)}]{2006A&A...453.1129P}
Paardekooper, S.-J. \& Mellema, G. 2006, A\&A, 453, pp.1129

\bibitem[{{Raymond} {et~al.}(2018){Raymond}, {Boulet}, {Izidoro}, {Esteves}, \&
  {Bitsch}}]{2018MNRAS.479L..81R}
{Raymond}, S.~N., {Boulet}, T., {Izidoro}, A., {Esteves}, L., \& {Bitsch}, B.
  2018, \mnras, 479, L81

\bibitem[{{Raymond} \& {Izidoro}(2017)}]{2017Icar..297..134R}
{Raymond}, S.~N. \& {Izidoro}, A. 2017, \icarus, 297, 134

\bibitem[{Santos {et~al.}(2004)Santos, Israelia, \&
  Mayor}]{2004A&A...415.1153S}
Santos, N.~C., Israelia, G., \& Mayor, M. 2004, A\&A, 415, p.1153

\bibitem[{{Savvidou} \& {Bitsch}(2021)}]{2021arXiv210311995S}
{Savvidou}, S. \& {Bitsch}, B. 2021, arXiv e-prints, arXiv:2103.11995

\bibitem[{{Savvidou} {et~al.}(2020){Savvidou}, {Bitsch}, \&
  {Lambrechts}}]{2020arXiv200514097S}
{Savvidou}, S., {Bitsch}, B., \& {Lambrechts}, M. 2020, arXiv e-prints,
  arXiv:2005.14097

\bibitem[{Schneider \& Bitsch(2021)}]{SchneiderBitsch}
Schneider, A. \& Bitsch, B. 2021, A\&A, in review

\bibitem[{{Schoonenberg} {et~al.}(2019){Schoonenberg}, {Liu}, {Ormel}, \&
  {Dorn}}]{2019A&A...627A.149S}
{Schoonenberg}, D., {Liu}, B., {Ormel}, C.~W., \& {Dorn}, C. 2019, \aap, 627,
  A149

\bibitem[{{Selsis} {et~al.}(2007){Selsis}, {Chazelas}, {Bord{\'e}}, {Ollivier},
  {Brachet}, {Decaudin}, {Bouchy}, {Ehrenreich}, {Grie{\ss}meier}, {Lammer},
  {Sotin}, {Grasset}, {Moutou}, {Barge}, {Deleuil}, {Mawet}, {Despois},
  {Kasting}, \& {L{\'e}ger}}]{2007Icar..191..453S}
{Selsis}, F., {Chazelas}, B., {Bord{\'e}}, P., {et~al.} 2007, \icarus, 191, 453

\bibitem[{{Shakura} \& {Sunyaev}(1973)}]{1973A&A....24..337S}
{Shakura}, N.~I. \& {Sunyaev}, R.~A. 1973, \aap, 24, 337

\bibitem[{Suzuki {et~al.}(2016)Suzuki, Ogihara, Morbidelli, Crida, \&
  Guillot}]{2016arXiv160900437S}
Suzuki, T.~K., Ogihara, M., Morbidelli, A., Crida, A., \& Guillot, T. 2016,
  A\&A, 596, id.A74

\bibitem[{{Terquem} \& {Papaloizou}(2007)}]{2007ApJ...654.1110T}
{Terquem}, C. \& {Papaloizou}, J. C.~B. 2007, \apj, 654, 1110

\bibitem[{{Tsiaras} {et~al.}(2019){Tsiaras}, {Waldmann}, {Tinetti}, {Tennyson},
  \& {Yurchenko}}]{2019NatAs...3.1086T}
{Tsiaras}, A., {Waldmann}, I.~P., {Tinetti}, G., {Tennyson}, J., \&
  {Yurchenko}, S.~N. 2019, Nature Astronomy, 3, 1086

\bibitem[{{Venturini} {et~al.}(2020){Venturini}, {Guilera}, {Haldemann},
  {Ronco}, \& {Mordasini}}]{2020A&A...643L...1V}
{Venturini}, J., {Guilera}, O.~M., {Haldemann}, J., {Ronco}, M.~P., \&
  {Mordasini}, C. 2020, \aap, 643, L1

\bibitem[{{Weber} {et~al.}(2018){Weber}, {Ben\'{\i}tez-Llambay}, {Gressel},
  {Krapp}, \& {Pessah}}]{2018ApJ...854..153W}
{Weber}, P., {Ben\'{\i}tez-Llambay}, P., {Gressel}, O., {Krapp}, L., \&
  {Pessah}, M.~E. 2018, \apj, 854, 153

\bibitem[{{Weidenschilling}(1977)}]{1977Ap&SS..51..153W}
{Weidenschilling}, S.~J. 1977, \apss, 51, 153

\bibitem[{{Zeng} {et~al.}(2019){Zeng}, {Jacobsen}, {Sasselov}, {Petaev},
  {Vanderburg}, {Lopez-Morales}, {Perez-Mercader}, {Mattsson}, {Li}, {Heising},
  {Bonomo}, {Damasso}, {Berger}, {Cao}, {Levi}, \&
  {Wordsworth}}]{2019PNAS..116.9723Z}
{Zeng}, L., {Jacobsen}, S.~B., {Sasselov}, D.~D., {et~al.} 2019, Proceedings of
  the National Academy of Science, 116, 9723

\bibitem[{{Zhu} {et~al.}(2012){Zhu}, {Nelson}, {Dong}, {Espaillat}, \&
  {Hartmann}}]{2012ApJ...755....6Z}
{Zhu}, Z., {Nelson}, R.~P., {Dong}, R., {Espaillat}, C., \& {Hartmann}, L.
  2012, \apj, 755, 6

\end{thebibliography}

\appendix

\section{Systems with inner small planets and gas giants}
\label{ap:systems}

We list here the planetary systems with transiting inner small planets (Table~\ref{tab:SE}) and their outer gas giant companions (Table~\ref{tab:CJ}), which are shown in Fig.~\ref{fig:MRCJ}. The data were taken from the NASA exoplanet archive\footnote{https://exoplanetarchive.ipac.caltech.edu/}.

We also list the transmission spectroscopy metric (TSM) value, as specified in \citet{2018PASP..130k4401K}. The TSM metric is proportional to the expected transmission spectroscopy S/N, based on the strength of spectral features ($\propto R_{\rm P} H / R_\star$ where $H$ is the atmospheric scale height) and the brightness of the host star, assuming cloud-free atmospheres (their Eq. 1). We mark in bold the systems that have TSM values above the thresholds suggested by \citet{2018PASP..130k4401K}. In addition, these three planets have errors in their measured masses below 20\% of their total mass, in line with the recommendations by \citet{2019ApJ...885L..25B} for atmospheric follow-up observations, making them optimal targets for testing our hypothesis.

\begin{table*}
\centering
\begin{tabular}{c|c|c|c|c|c|c|c}
\hline
Name    &       mass [$\rm M_{\rm E}$] &        radius [$\rm R_{\rm E}$]&       $\bar\rho$ [g/cm$^3$] & a [AU]&    period [days]&  Mstar [$M_\odot$] & TSM \\ \hline \hline
55-Cnc-e&       7.81 $\pm$ 0.56 &       2.08 $\pm$ 0.16&        4.78&   0.01543&        0.736&  1.015 & {\bf 354} \\
K-97b   &       3.51 $\pm$ 1.9 &        1.48 $\pm$ 0.13&        5.97&   0.0361&         2.586&   0.94 & 1.2 \\
K-93b   &       4.02 $\pm$ 0.68 &       1.48 $\pm$ 0.00006&     6.83&   0.0534&         4.7267&  0.91 & 3.6 \\ 
K-68b   &       7.65 $\pm$ 1.32 &       2.31 $\pm$ 0.1& 3.42&   0.0617&     5.498&       1.079 & 25.8\\
K-68c   &       2.04 $\pm$ 1.72 &       0.958 $\pm$ 0.05&       12.79&  0.0906&     9.60&        1.079 & 0.87\\
K-48b   &       3.94 $\pm$ 2.10 &       1.88 $\pm$ 0.1& 3.26&   0.0532&         4.778&   0.88  & 11.5\\
K-48c   &       14.611 $\pm$ 2.3 &      2.71 $\pm$ 0.14&        4.05&   0.0851&         9.67&    0.88  & 7.38\\
K-48d   &       7.9 $\pm$ 4.6 & 2.04 $\pm$ 0.11&        5.12&   0.2298&         42.89&   0.88  & 3.55\\
K-407b  &       $<$3.2 &        1.07 $\pm$ 0.02&        14.4&   0.0150&         0.6693&  1.0   & 0.82 \\
K-94b   &       10.841 $\pm$ 1.4 &      3.51 $\pm$ 0.15&        1.38&   0.0337&         2.508&   0.81 & 57 \\
$\pi$-Men-c&    4.52 $\pm$ 0.81 &       2.06 $\pm$ 0.03&        2.85& 0.067&        6.268&       1.094 & {\bf 252}\\     
WASP-47e&       6.83 $\pm$ 0.66 &       1.81 $\pm$ 0.27&        6.34&   0.0173&     0.789&       1.10  & 14.2\\
WASP-47d&       13.1 $\pm$ 1.5 &        3.576 $\pm$ 0.046&      1.58&   0.0886&     9.03&        1.10  & 25.5\\
K-454b  &       6.05 $\pm$ 1.51 &       2.4 $\pm$ 0.2&  2.38&   0.0954&     10.57&       1.028 & 19.4\\
K-289b  &       7.3 $\pm$ 6.8 & 2.15 $\pm$ 0.1& 4.05&   0.21&       34.545&     1.08  & 2.84\\
K-289d  &       4.1 $\pm$ 0.9 & 2.68 $\pm$ 0.17&        1.17&   0.33&       66.06&       1.08  & 7.94 \\
HD-86226c &   7.25 $\pm$ 1.15 & 2.16 $\pm$ 0.08&   3.97&   0.049&      3.98&  1.019 & 88.7\\
HAT-P-11b & 23.4 $\pm$ 1.5 & 4.36 $\pm$ 0.06&     1.59&      0.053&      4.88 &  0.81 & {\bf 192} \\
K-88b   &   9.5 $\pm$ 1.10 &    3.44 $\pm$ 0.075&   1.29&   0.09&       10.916&  0.99 & 23.3\\  
\hline
\end{tabular}
\caption[Super-Earths]{Transiting sub-Neptune systems with outer giant planets (Table~\ref{tab:CJ}). Data for the systems were taken from \url{https://exoplanetarchive.ipac.caltech.edu}. The mean density, $\bar \rho$, has been calculated from the mass and radius. As a reference, the Earth's mean density is 5.51g/cm$^3$.}
\label{tab:SE}
\end{table*}

\begin{table*}
\centering
\begin{tabular}{c|c|c|c|c|c|c}
\hline
Name    &       mass [$\rm M_{\rm E}$] &        radius [$\rm R_{\rm E}$]&       a [AU]&   Mstar [$M_\odot$] & distance [pc] & [Fe/H] \\ \hline \hline
55-Cnc-b&       255 $\pm$ 2.9&  -&      0.1133 $\pm$ 0.0006&    1.015 & 12.34 & 0.35 \\
55-Cnc-c&   51.2  $\pm$ 1.3&    -&  0.237  $\pm$ 0.0013&  1.015 & 12.34 & 0.35 \\
55-Cnc-f&   47.8  $\pm$ 2.4&    -&  0.7733 $\pm$ 0.0043&   1.015 & 12.34 & 0.35 \\
55-Cnc-d&  992 $\pm$ 33&    -&  5.446 $\pm$ 0.074&   1.015 & 12.34 & 0.35 \\
K-97c   &       343 &   -&      1.636 &  0.94 & 405 & -0.2 \\
K-93c   &       2700 &  -&      4.536 &    0.91 & 96.7 & -0.18 \\
K-68d   &       267 $\pm$ 16&   -&      1.40 $\pm$ 0.03&        1.079 & 135 & 0.12 \\
K-48e   &       657 $\pm$ 25&   -&      1.834 $\pm$ 0.03&       0.88 & 313 & 0.3 \\
K-407c  &       3000 &  -&      4.07 &          1.0 & 338 & 0.34\\
K-94c   &       3125 $\pm$ 200& -&      1.58 $\pm$ 0.03&        0.81 & 223 & 0.34 \\
$\pi$-Men-b&    3070 $\pm$ 64&  -&      3.10 $\pm$ 0.03 &               1.094 & 18.37 & 0.08\\
WASP-47b&       363 $\pm$ 7.3&  13.1&   0.052 $\pm$ 0.011&      1.10 & 200 & 0.36 \\
WASP-47c&       481 $\pm$ 220&  -&      1.41 $\pm$ 0.30&        1.10 & 200 & 0.36 \\
K-454c  &       1418 $\pm$ 38&  -&      1.286 $\pm$ 0.001&      1.028 & 200 & 0.27 \\
K-289c  &       131.9 $\pm$ 17& 11.59&  0.51 $\pm$ 0.03&        1.08 & 700 & 0.33 \\
HD-86226b & 292 $\pm$ 32&  -&   2.73 $\pm$ 0.06&   1.019 & 45.57 & 0.02\\
HAT-P-11c & 731 $\pm$ 200&  -&   4.36 $\pm$ 0.2&   0.81 &  37.89 & 0.32 \\
K-88c   &   214 $\pm$ 5.2&  -&   0.1529 $\pm$ 0.001&    0.99&   385 & 0.27 \\
K-88d   &   965 $\pm$ 44&  -&   2.45 $\pm$ 0.01&  0.99& 385 & 0.27 \\
\hline
\end{tabular}
\caption[Cold Jupiters]{Cold Jupiters with inner sub-Neptunes, which are listed in Table~\ref{tab:SE}. Data for the systems were taken from exoplanet.eu and the open exoplanet catalog. Planets with no indicated radius are not transiting. These data were taken from \url{https://exoplanetarchive.ipac.caltech.edu}. Some planets do not contain errors in the database. [Fe/H] indicates the host star metallicity.}
\label{tab:CJ}
\end{table*}

\section{Water content in the inner disk}
\label{ap:water}

In order to understand how inward drifting pebbles that evaporate at the water ice line influence the water content of the gas in the inner disk, we employed a simple pebble drift and evaporation model that includes a prescription of a pressure bump corresponding to a growing planet. The details of this setup will be described in an upcoming paper \citep{SchneiderBitsch}; here we just discuss the underlying mechanism.

The basis of our pebble growth and drift model is based on the model of \citet{2012A&A...539A.148B}, which we extended with pebble evaporation and condensation at ice lines $r_{\mathrm{ice,Y}}$. This adds a source term to the evolution of the solid density given by
\begin{equation}\label{eq:sink_ice}
        \dot\Sigma_Y = 
        \begin{cases}
                \dot\Sigma^{\mathrm{evap}}_\mathrm{Y} & r<r_{\mathrm{ice,Y}}\\ 
                \dot\Sigma^{\mathrm{cond}}_\mathrm{Y}    & r\geq r_{\mathrm{ice,Y}}\\ 
        \end{cases}
.\end{equation}
Here $\dot\Sigma^{\mathrm{evap}}_{\rm Y}$ and $\dot\Sigma^{\mathrm{cond}}_{\rm Y}$ are the evaporation and condensation source terms of species Y (e.g., water) for the two transport equations of gas and dust. For the condensation term we assumed that gas can only condensate by sticking onto the surface (with efficiency $\epsilon_p=0.5$) of existent solids. The condensation term is then given by: 
\begin{equation}
        \dot\Sigma^{\mathrm{cond}}_\mathrm{Y} = 
\frac{3\epsilon_p}{2\pi\rho_\bullet}\Sigma_{\mathrm{g,Y}}\left(\frac{\Sigma_\mathrm{dust}}{a_\mathrm{dust}}+\frac{\Sigma_\mathrm{peb}}{a_\mathrm{peb}}\right)\Omega_\mathrm{k}\sqrt{\frac{\mu}{\mu_\mathrm{Y}}},
\end{equation}
where $\mu_\mathrm{Y}$ is the mass (in proton masses) of a molecule of species Y. Here $a_\mathrm{dust}$ and $a_\mathrm{peb}$ are the particle sizes of the small and large dust distribution with corresponding surface densities $\Sigma_{\rm dust}$ and $\Sigma_{\rm peb}$, respectively, $\rho_\bullet$ denotes the pebble density, which we set to 1.6 g/cm$^3$, and $\Omega_{\rm K}$ denotes the Keplerian frequency.

For the evaporation term we assumed that the flux of solids that drifts through the evaporation line evaporates into gas within 0.1 AU:
\begin{equation}\label{eq:sink_ice_evap}
        \dot\Sigma^{\mathrm{evap}}_Y = \frac{\Sigma_\mathrm{Z,Y}\cdot \bar u_{\mathrm{Z}}}{{0.1}{\rm AU}}
.\end{equation}
Here $\Sigma_{\rm Z,Y}$ is the solid surface density of species Y and $\bar{u}_{\rm Z}$ denotes the mean radial velocity of the solid dust grains.

\begin{figure*}
 \centering
 \includegraphics[scale=0.45]{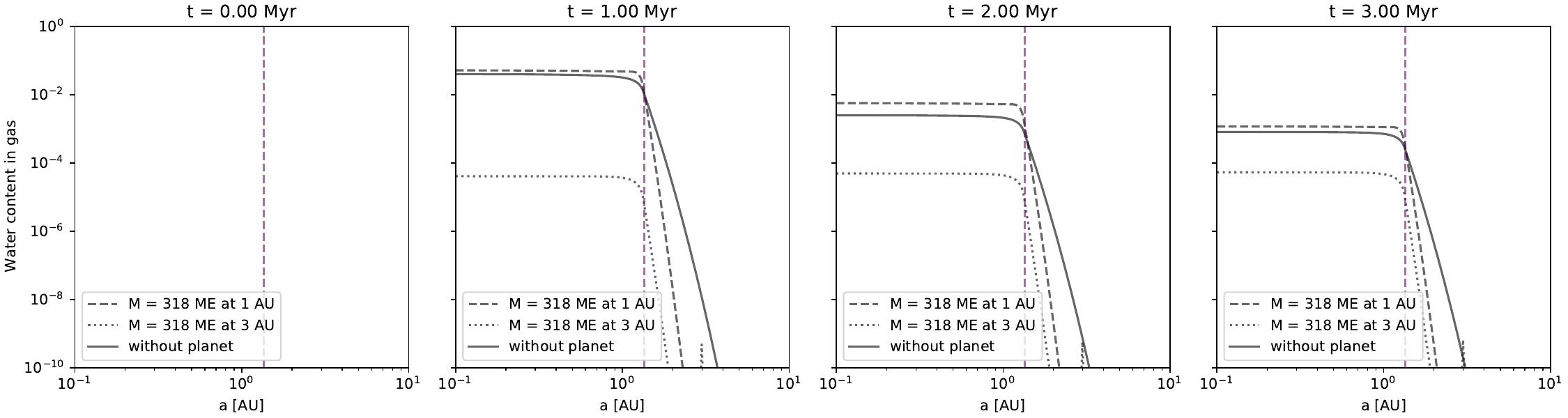}  
 \includegraphics[scale=0.45]{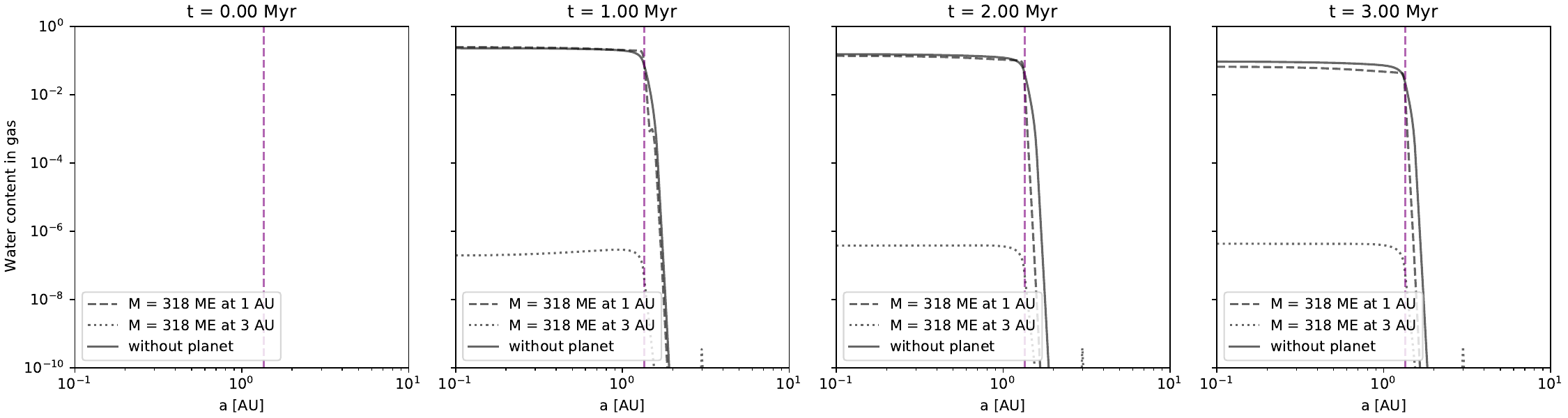} 
 \caption{Water content in the gas phase for our three different scenarios (Fig.~\ref{fig:Cartoon}) as a function of time, where the giant planet is placed inside (dashed lines, scenario A) or outside (dotted lines, scenario B) the water ice line or where there is no giant planet (solid line, scenario C). The panels in the top row show a disk evolution with $\alpha=0.001$, while the bottom row shows a disk evolution with $\alpha=10^{-4}$. The 1 Jupiter mass planet is placed in the disk at the beginning of the simulation. The vertical line shows the position of the water ice line, where some water vapor can also exist outside the ice line due to the outward diffusion of water vapor and not immediate re-condensation. It is clear that a planet positioned outside the water ice line is efficient in blocking pebbles, resulting in a low water fraction of the gas phase in the inner disk. The pebble flux diminishes over time, reducing the water vapor in the inner disk for all cases.
   \label{fig:discwater}
   }
\end{figure*}

In Fig.~\ref{fig:discwater} we display the water content in the gas for our three different scenarios and for $\alpha=0.001$ and $\alpha=10^{-4}$, where we have placed a 1 Jupiter mass planet in the disk at the beginning of the simulation. The gap depth was calculated via gravitational gap clearing \citep{2006Icar..181..587C}. The aspect ratio in the inner disk is relatively small, and therefore planets of already a few tens of Earth masses can reach the pebble isolation mass and block inward drifting pebbles (e.g., \citealt{2019A&A...623A..88B}). Depending on the relative position of the giant planet to the water ice line, the inner disk is water rich or water poor (see Fig.~\ref{fig:Cartoon}). Clearly, giant planets inside the water ice line do not prevent a large water vapor accumulation in the inner disk (scenario B).

Furthermore, the water content of the gaseous component of the inner disk is at most $\approx$20\% (at low $\alpha$), indicating that sub-Neptunes that form in scenario B (Fig.~\ref{fig:Cartoon}) should have maximum have a water content on that order in their atmosphere. The exact water content of the inner disk depends crucially on the disk's viscosity, which sets the pebble sizes as well as the diffusion of gas and pebbles through the pressure bump. A lower viscosity results in larger pebbles, which drift inward faster \citep{2012A&A...539A.148B} and can thus enrich the inner disk with more water vapor in a shorter time (top row in Fig.~\ref{fig:discwater}). Additionally, a lower viscosity diffuses the vapor-rich gas inward more slowly, enhancing this effect and allowing a larger water vapor content to remain for longer times compared to higher viscosities. On the other hand, the lower viscosity prevents pebble diffusion through the planetary gap (resulting in a water-poor inner disk) if the giant planet is outside the water ice line (Fig.~\ref{fig:discwater}), as proposed in scenario A (Fig.~\ref{fig:Cartoon}). However, if the water content of an observed sub-Neptune is significantly higher than that of the disk's gas phase, it indicates that the planet should have formed at the water ice line and then migrated inward \citep{2019A&A...624A.109B, 2019arXiv190208772I, 2020A&A...643L...1V}, as proposed in scenario C.

\end{document}